\documentclass[]{iopart}

\usepackage{graphicx}
\usepackage{dcolumn}
\usepackage{bm}

\begin{document}


\title{Direct observation of Fe spin reorientation in single crystalline YbFe$_6$Ge$_6$}

\author{M. A. Avila$^1$, T. Takabatake$^1$, Y. Takahashi$^2$, S. L. Bud'ko$^3$, P. C. Canfield$^3$}
\address{%
$^1$ Department of Quantum Matter, ADSM, Hiroshima University,
Higashi-Hiroshima 739-8530
}%
\address{%
$^2$ Deptartment of Earth and Planetary Systems Sciences,
Hiroshima University, Higashi-Hiroshima 739-8526
}%
\address{%
$^3$ Ames Laboratory and Department of Physics and Astronomy, Iowa
State University, Ames, IA 50011
}%

\date{\today}

\begin{abstract}
We have grown single crystals of YbFe$_6$Ge$_6$ and LuFe$_6$Ge$_6$
and characterized their anisotropic behaviour through low field
magnetic susceptibility, field-dependent magnetization,
resistivity and heat capacity measurements. The Yb$^{+3}$ valency
is confirmed by L$_{III}$ XANES measurements. YbFe$_6$Ge$_6$
crystals exhibit a field-dependent, sudden reorientation of the Fe
spins at about 63~K, a unique effect in the RFe$_6$Ge$_6$ family
(R~=~rare earths) where the Fe ions order anti-ferromagnetically
with Ne\'{e}l temperatures above 450~K and the R ions' magnetism
appears to behave independently. The possible origins of this
unusual behaviour of the ordered Fe moments in this compound are
discussed.

\end{abstract}



\section{Introduction}

The combination of localized $4f$-electron magnetism with
delocalized $3d$-electron or band magnetism can often provide both
interesting physical phenomena from the academic perspective, and
potentially useful effects from the application perspective, thus
making it a continuous topic of interest in materials science and
magnetism. To study the coexistence and relationships between
these different types of magnetic entities in the same compound,
one will usually focus on binary or ternary intermetallics
containing both rare-earth elements (Ce-Yb) and $3d$-shell
transition metals Mn-Ni.

\begin{figure}[htb]
\begin{center}
\includegraphics[width=99mm]{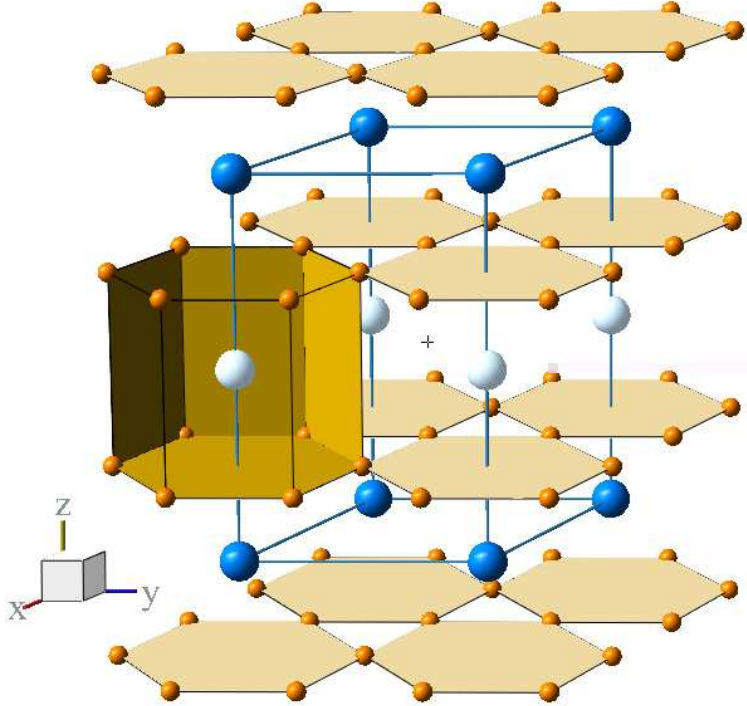}
\caption{\label{structure} Schematic representation of the
HfFe$_6$Ge$_6$-type structure ($P6/mmm$ space group). Small
spheres are the Fe $6i$ sites, large light spheres are the Hf
atom's ideally occupied $1b$ sites and large dark spheres are the
ideally unoccupied $1a$ sites. The three Ge sites have been left
out for clarity: they form hexagonal bipyramid Ge$_8$ cages around
unoccupied Hf sites, or a simple co-planar hexagonal grid around
occupied Hf sites. The 8 unoccupied Hf sites delimit the
structure's unit cell.}
\end{center}
\end{figure}

The RT$_6$X$_6$ (R = rare earths; T = Mn, Fe, Co; X = Ge, Sn) is
one such family of intermetallics that has been known for several
decades, and many new members of the family have been synthesized
and explored extensively over the past decade or so
 \cite{vent92a,wang94a}. Its formation can be viewed as an
insertion of R atoms into the layered, hexagonal FeGe-type binary
structure (Figure~\ref{structure}). In the ideal arrangement, R
atoms alternate between complete occupation of an interstitial
layer between two hexagonal Fe grids, and no occupation of the
adjacent interstitial layers, so that the unit cell is doubled
along the c-axis with respect to the original FeGe structure. This
is known as the HfFe$_6$Ge$_6$-type structure, adopted by most of
the RMn$_6$Ge$_6$ compounds, but only achievable by the smallest
of the R ions such as Lu and Yb in the RFe$_6$Ge$_6$ series
 \cite{vent92a,chab83a}. In actual samples, all interstitial layers
tend to become partially occupied, sometimes in a disordered
version of the alternating layer structure for the smaller R ions
\cite{papa98a}, and sometimes in organized manners which lead to
different symmetries and sizes of the unit cell for larger R ions
\cite{vent92a}. Thus, the compounds in this family are
intrinsically prone to disorder as one might expect, which often
leads to batch and thermal history dependencies of a sample's
structure and properties \cite{wang94a,zaha98a,zaha99a}.

Whatever the case, a crucial feature of these structures is that
the R atoms always position themselves in the center of a
hexagonal prism formed by twelve T atoms (highlighted in
figure~\ref{structure}). In most of the RMn$_6$Sn$_6$ compounds,
the Mn ions order ferromagnetically at temperatures between 350
and 450~K. The 12 Mn moments in the hexagonal prism and the R
magnetic moment in its center interact strongly, with negative
exchange coefficient, and the R sub-lattice was found to order
simultaneously with the Mn sub-lattice, forming a ferrimagnetic
structure in the material \cite{mala99a,clat99a}. In the
RMn$_6$Ge$_6$ series, a set of complex ferro-, ferri- and
antiferromagnetic arrangements including multiple wave-vectors and
transitions have been observed
 \cite{vent92a,idri94a,ros96a,vent95a,brab93a,papa95a,papa95b}.

In the RFe$_6$Ge$_6$ series, the Fe ions retain much of their
magnetic behaviour from the parent FeGe compound. At temperatures
varying between 450 and 480 K \cite{vent92a,wang94a}, they order
ferromagnetically and axially within a single hexagonal layer, but
neighboring Fe layers order antiferromagnetically, and thus the
material assumes a simple $+ - + -$ stacking of spins along the
$c$-axis. As a consequence, the net local field at the R site due
to its twelve Fe neighbors of the hexagonal prism is null, and it
has been found that the rare-earth sub-lattice does not order
together with the Fe sub-lattice, but rather behaves quite
independently \cite{ryan96a} - ordering at much lower temperatures
in some cases, or remaining paramagnetic down to the lowest
measured temperatures in others. These cases were also observed in
a few of the manganese based compounds where antiferromagnetic
ordering of the Mn ions occurs \cite{clat99a}.

YbFe$_6$Ge$_6$ has been reported to exhibit antiferromagnetism of
the kind described above \cite{vent92a,ryan96a}. Lattice
parameters indicated that the Yb ion should be in or close to its
+3 state (and therefore magnetic) but remains paramagnetic down to
1.5 K. However, a neutron diffraction and M\"{o}ssbauer study of
polycrystalline samples by Mazet and Malaman \cite{maz00a} showed
evidence of a spin reorientation effect where at least part of the
Fe spins in their samples deviated from the $c$-axis below about
85 K, assuming a direction close to the basal plane. Such
reorientation processes of the ordered 3$d$ spins at intermediate
temperatures have been well documented in some of the
RMn$_6$Sn$_6$ compounds \cite{maz01a}, and in this series it has
been proposed that the effect results from the interplay between
two competing energies: the Mn ferromagnetic interaction which
dominates the high temperature region and favors an easy-plane
arrangement, and the rare-earths' crystal field anisotropy which
becomes dominant at lower temperatures, forcing the R spins
towards the $c$-axis and the Mn spins to shift simultaneously, due
to the strong R-Mn interaction \cite{maz01a}.

However, such a drastic spin reorientation effect is not to be
expected in the ``independent sub-lattices'' scenario of the
RFe$_6$Ge$_6$ series, and indeed it has not been found in other
members of the series. We were thus motivated by this puzzling
behaviour of YbFe$_6$Ge$_6$ to address the issue by growing single
crystals in order to study the material's anisotropic behaviour
directly. In this work we present further evidence of this spin
reorientation process in our single crystals, and discuss the
possible mechanisms that may be responsible for its unique
occurrence in the RFe$_6$Ge$_6$ series.

\section{Experimental Details}

Due to the relevance of thermal history previously mentioned, we
wish to provide a description of our growth procedures in detail.
Single crystals of YbFe$_6$Ge$_6$ and LuFe$_6$Ge$_6$ were grown
using Sn as a fourth-element flux \cite{can92a}.

\begin{figure}[htb]
\begin{center}
\includegraphics[angle=0,width=99mm]{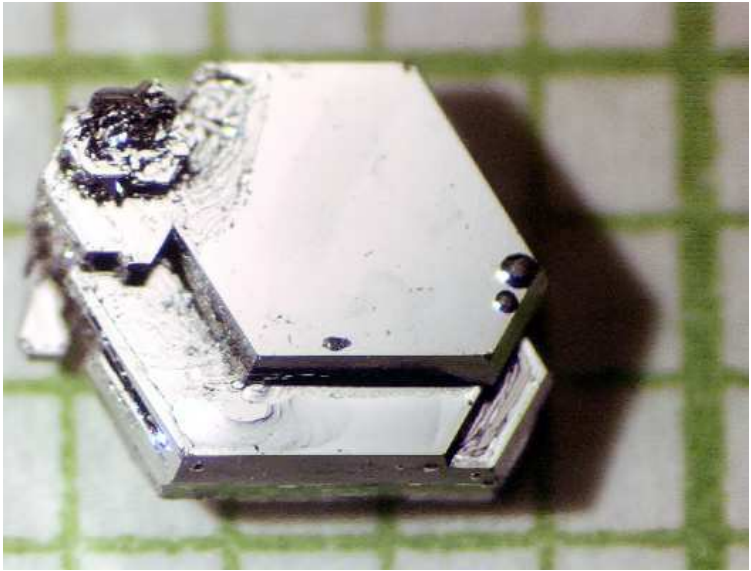}
\caption{\label{crystal} As-grown single crystal of LuFe$_6$Ge$_6$
on a millimeter paper. The rounded droplet-like features located
at 4:00 and 10:00 are solidified Sn flux.}
\end{center}
\end{figure}

For LuFe$_6$Ge$_6$, thick hexagonal crystal plates weighing up to
16.5 mg (figure~\ref{crystal}) were obtained by mixing high-purity
starting elements (99.95\% or better) with a Lu:Fe:Ge:Sn
proportion of 1:6:6:20 in an alumina crucible, which was then
sealed under partial argon atmosphere inside a quartz ampoule. The
ampoule was placed in a box furnace, heated to 1200~$^\circ$C and
maintained for 2 hours, then cooled over several days to
500~$^\circ$C, at which point the ampoule was quickly removed from
the furnace and the molten Sn flux was separated by decanting. The
decanted flux as well as most of the as-grown crystals reacted to
the presence of magnets at room temperature, indicating that a
ferromagnetic second phase also precipitated out of the Sn flux
during the cooling process (probably an Fe-rich alloy since Fe
isn't very soluble in Sn) so the crystals were placed in a bath of
50\% HCl in water for several days to remove any surface-attached
second phase (the RFe$_6$Ge$_6$ crystals are stable even in
concentrated HCl). Any crystals that still reacted to magnets
after this treatment were discarded since the remaining second
phase is trapped inside the crystal and is unreachable by the
acid.

YbFe$_6$Ge$_6$ single crystals were somewhat more difficult to
grow. Applying the same procedure described above results in very
small and thin (sub-milligram) hexagonal plates. However, this has
proved to be the only phase that crystallizes over a wide range of
Yb:Fe:Ge starting proportions dissolved in Sn, so this feature
could be explored to our advantage. We were able to increase the
crystal sizes up to 4.5 mg in a growth where an iron-deficient
starting mixture was used (Yb:Fe:Ge:Sn proportion of 1:1:3:20)
directly in the quartz ampoule, a soaking time of 10 hours at
1200~$^\circ$C and cooling to 500~$^\circ$C over 4 days. Some
small volume crystals grew as thin, elongated plates exceeding 2
mm in length along an $a$-axis, which were ideal for the 4-probe
resistivity measurements. We could also grow elongated rods along
the $c$-axis instead of plates in a batch using a 1:3:6:40
starting proportion and cooling over 4 days from 800~$^\circ$C to
600~$^\circ$C. None of these crystals reacted to magnets at room
temperature, but were still cleaned in HCl baths to remove any
remaining flux droplets from the surface.

\begin{figure}[htb]
\begin{center}
\includegraphics[angle=0,width=99mm]{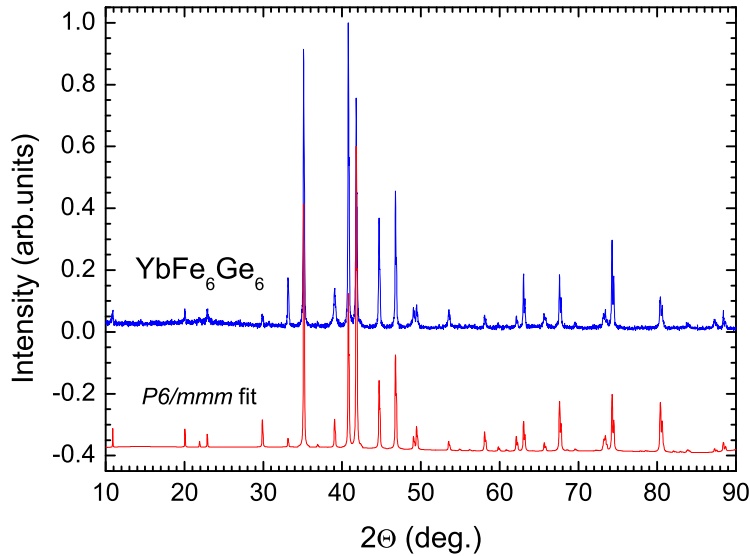}
\caption{\label{ybxray} Powder X-ray diffraction pattern of
crushed YbFe$_6$Ge$_6$ crystals and corresponding refinement.}
\end{center}
\end{figure}

The phases were checked by powder x-ray diffraction on crushed
crystals. Figure~\ref{ybxray} shows the measured pattern of
YbFe$_6$Ge$_6$ and its corresponding $P6/mmm$ refinement, which
results in cell parameters $a=5.102(1)${~\AA} and
$c=8.099(1)${~\AA}, in good agreement with the available
crystallographic data \cite{vent92a}.

Resistivity was measured on a home-made dc system and on Quantum
Design PPMS systems. The latter were also used to perform heat
capacity experiments. DC magnetization measurements were performed
on Quantum Design MPMS magnetometers. Ytterbium L$_{III}$-edge
XANES was measured at beamline BL01B1 of SPring-8, Hyogo, Japan. A
Si(111) double-crystal monochromator with two mirrors was used to
obtain the incident X-rays. XANES spectra were measured in
fluorescence mode using Lytle detector for the powdered samples at
Yb 1.0 wt.\% diluted by boron nitride. Energy calibration was
conducted by defining the peak energy of the white line for
Yb$_2$O$_3$ at 8.9245 keV.  The absorption of the spectra was
normalized to the average absorption between 8.98 and 9.00 keV.

\section{Experimental Results}

For comparative purposes, it is useful to initially focus on the
anisotropic behaviour of LuFe$_6$Ge$_6$, where the closed
electronic $4f$ shell of the Lu$^{+3}$ ion does not contribute to
the magnetic behaviour. Previous studies on polycrystalline
samples (and several others with non-magnetic R elements) have
shown AFM transitions above 450~K \cite{maz01a}, but the simple AF
arrangement of the Fe spins was found to become unstable at lower
temperatures, such that at least part of the Fe spins no longer
stay completely aligned with the $c$-axis \cite{maz01a,maz00b}.
This so-called ``spin-disorientation'' effect \cite{maz01a}
manifests in magnetic measurements as a return to an increasing
susceptibility regime as the temperature is lowered, contrary to
the expected behaviour of a material in the AFM state.

\begin{figure}[htb]
\begin{center}
\includegraphics[angle=0,width=99mm]{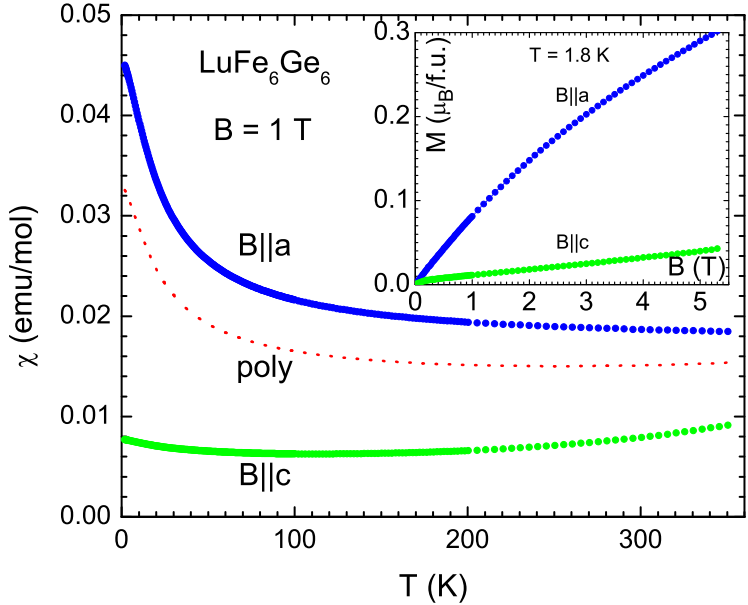}
\caption{\label{lumag} Anisotropic susceptibility at $B=1$~T of
LuFe$_6$Ge$_6$ in the AFM state. Inset: magnetization isotherms up
to 5~T at $T=1.8$~K.}
\end{center}
\end{figure}

Although our experimental system does not reach the temperature
range of the AFM transition, the measurements we have made below
350 K on a LuFe$_6$Ge$_6$ single crystal (after cleaning) supports
this scenario. Figure~\ref{lumag} shows that the magnetic
susceptibility within the AFM state is anisotropic and much larger
for B$\parallel$a than for B$\parallel$c. The B$\parallel$c curve
shows a broad local minimum at about 120 K, close to the region
where deviations from axial arrangement are reported to begin
 \cite{maz01a}.

It is also worth noting that the lowest temperature susceptibility
is non-Curie-Weiss like and tends towards ending in a cusp or
saturation. The increase observed for B$\parallel$a is quite
strong and, by forcing a Curie-Weiss fit in the polycrystalline
average curve below 100~K, we find that it would require a
contamination of the order of 8\% Tb (for example) to produce a
comparable anisotropic increase, indicating that the observed
upturn is more likely attributed to intrinsic behaviour than to
impurities. The reversible magnetization curves shown in the inset
of figure~\ref{lumag} also reveal no significant ferromagnetic
component that could arise from impurity phases.

\begin{figure}[tb]
\begin{center}
\includegraphics[angle=0,width=99mm]{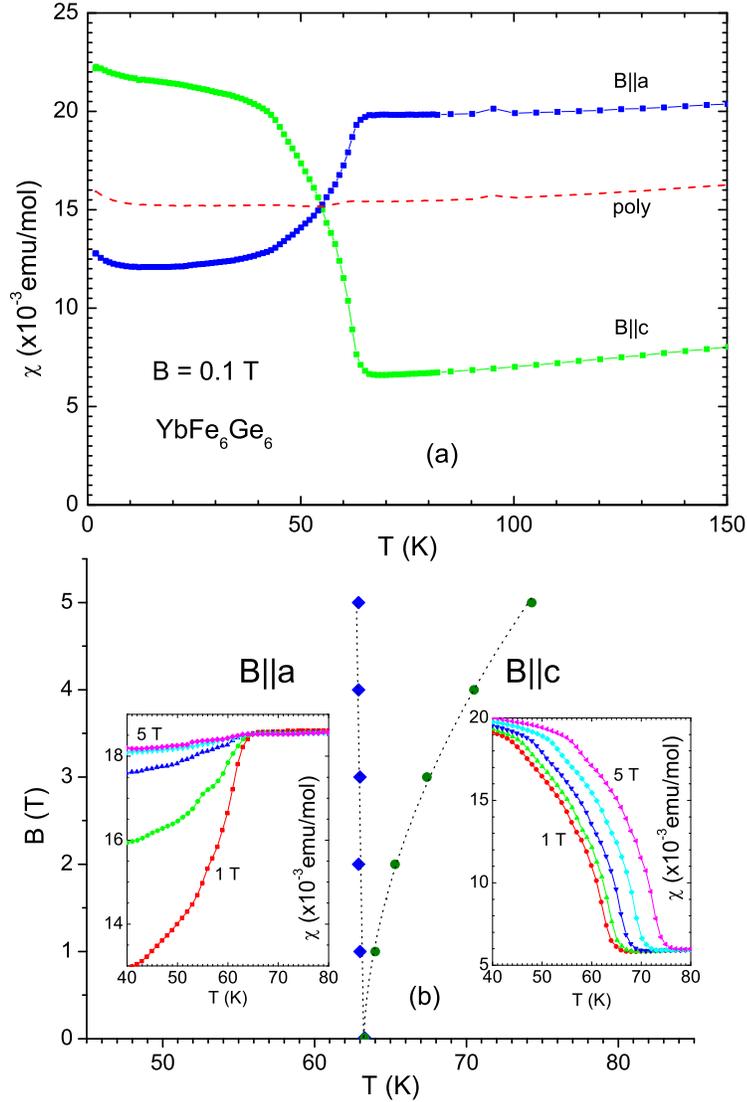}
\caption{\label{ybmag} Low-temperature magnetic behaviour of
YbFe$_6$Ge$_6$. (a) Anisotropic susceptibility at $B=0.1$~T,
showing the anomalous transition around 63~K. The dotted line is
the polycrystalline average. (b) Field dependence of the
transition temperature $T_{SR}$ for both orientations. The dotted
lines are guides to the eyes. The left and right insets show how
the transitions evolve under applied fields of 1, 2, 3, 4, and 5
T, for B$\parallel$a and B$\parallel$c respectively. }
\end{center}
\end{figure}

To our knowledge, no magnetization measurements on any type of
YbFe$_6$Ge$_6$ sample have been reported so far. The orientation
dependent measurements we have performed below 350 K on our single
crystals gave results generally similar to that of LuFe$_6$Ge$_6$
until about 63 K, at which point a sudden and drastic anomaly in
behaviour takes place. Figure~\ref{ybmag}a shows the curves for an
applied field of 0.1 T. The B$\parallel$c susceptibility increases
rapidly while the B$\parallel$a susceptibility decreases half as
much, in such a way that the material's anisotropy seems to be
almost completely inverted. The polycrystalline average (dotted
line) estimated from these two measurements as
$\chi_p=(\chi_c+2\chi_{a})/3$ remains smooth and virtually
unaffected, so low-field magnetization measurements on
polycrystals may indeed not reveal any hint that such a transition
is taking place. It is interesting to note that there is no clear
manifestation of Yb$^{+3}$ paramagnetism in these measurements.

The anomalous behaviour is field-sensitive and evolves quite
differently in each orientation. For B$\parallel$c (right inset of
figure~\ref{ybmag}b), higher fields favor the transition, moving
it to higher temperatures. Defining $T_{SR}$ as the point where
the mid-transition (highest-slope) extrapolation meets the
above-transition (baseline) extrapolation, we see $T_{SR}$
increasing from 64.0 K at 1 T to 74.3 K at 5 T. The curve shape,
however, remains essentially the same in this field interval. For
B$\parallel$a, by contrast, higher fields clearly act towards
suppressing the transition, both in terms of moving $T_{SR}$ to
slightly lower temperatures, and in terms of decreasing the
magnitude of the susceptibility change, until it has almost
vanished at 5 T. This field dependence is consistent with simple
energetics: high-fields stabilize the high magnetization phase and
suppress the low magnetization phase. Because of this difference,
polycrystalline average curves estimated for these high field
measurements are no longer smooth and featureless as those for low
fields, so the transition in magnetization should also become
quite evident in powders and sintered samples measured at high
enough fields.

The main graph in figure~\ref{ybmag}b shows a tentative phase
diagram for how $T_{SR}$ evolves with field in both orientations.
Measurements made at low fields showed that $T_{SR}$ in both
directions essentially coincide at 63.3 K for this sample.
However, earlier batches showed quite different values of $T_{SR}$
(as low as 44 K) and multiple transitions, pointing to the
relevance of disorder and thermal history in this system, which
will be discussed later.

\begin{figure}[htb]
\begin{center}
\includegraphics[angle=0,width=99mm]{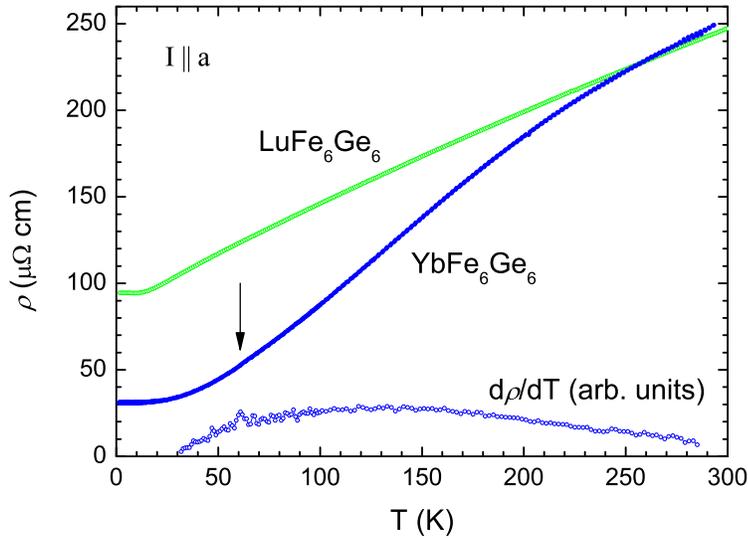}
\caption{\label{transp} In-plane electrical resistivity $\rho(T)$
of YbFe$_6$Ge$_6$ and LuFe$_6$Ge$_6$ below room temperature. The
derivative $d\rho/dT$ for the former is also shown.}
\end{center}
\end{figure}

Having documented the magnetic behaviour, we now present a few
other experimental techniques that provide further
characterization of the compound and information about the nature
of the observed transition. Figure~\ref{transp} shows resistivity
measurements performed on an $a$-axis elongated YbFe$_6$Ge$_6$
crystal. Room temperature resistivity is about 250~$\mu\Omega$~cm,
decreasing to 31~$\mu\Omega$~cm at the lowest measured temperature
and resulting in $RRR=\rho(300K)/\rho(0)=8.1$, a surprisingly high
value for a compound that is so prone to disorder. A measurement
on a LuFe$_6$Ge$_6$ crystal cut into bar shape is also shown,
although only qualitative comparison in transport properties can
be made.

Only a very subtle change in the resistivity behaviour (and thus
scattering regime) of YbFe$_6$Ge$_6$ was found in this
measurement, observable as a peak that barely rises above noise
level in the derivative $d\rho/dT$ (the arrow marks $T_{SR}$ as
observed by magnetization of the same sample). This minor change
gives qualitative support to the fact that the transition involves
a spin reorientation, and not a more fundamental structural or
magnetic phase transition in the sample, in which cases much more
pronounced changes in the scattering regime are usually expected.
Small features in resistivity like this were also observed in the
reorientation transitions of some RMn$_6$Ge$_6$ compounds
 \cite{dui99a}.

\begin{figure}[htb]
\begin{center}
\includegraphics[angle=0,width=99mm]{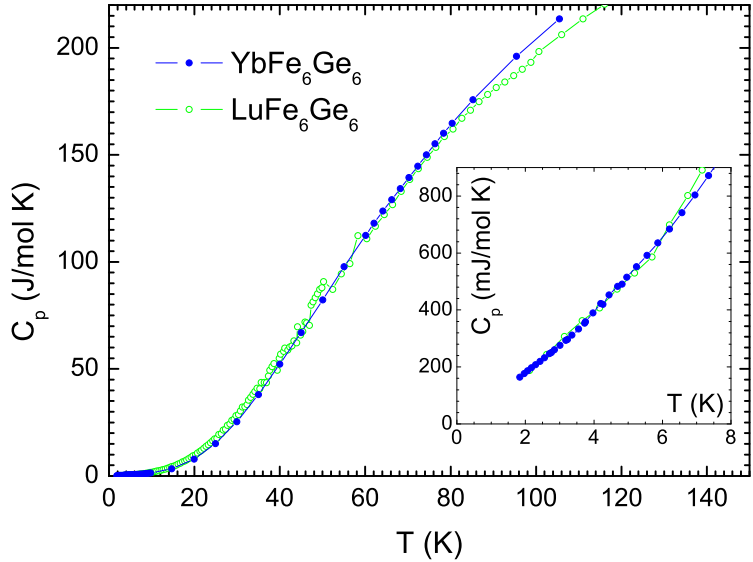}
\caption{\label{heat} Zero-field specific heat of YbFe$_6$Ge$_6$
and LuFe$_6$Ge$_6$ below 100~K. The inset shows the similarity in
behaviour persists down to the lowest measured temperature of
1.8~K.}
\end{center}
\end{figure}

Another useful measurement to help understand magnetic behaviours
and/or the nature of phase transitions is the
temperature-dependent heat capacity. In figure~\ref{heat} we show
the zero-field specific heat of both YbFe$_6$Ge$_6$ and
LuFe$_6$Ge$_6$. The behaviour of both compounds is very similar,
even at the lowest measured temperatures (inset), indicating that
no major or drastic changes in entropy take place in
YbFe$_6$Ge$_6$ in this temperature interval, compared to
LuFe$_6$Ge$_6$. It is possible though, that more careful
measurements in the region of interest could reveal some subtle
feature such as the one found in resistivity. The Sommerfeld
coefficient estimated by extrapolating $Cp/T vs. T^2$ to $T=0$ is
about 90~mJ/mol~K for both samples. The coincidence in behaviour
of both compounds indicates that any Yb sub-lattice magnetism
contributes very little to the overall entropy, and therefore the
behaviour is essentially dominated by the Fe magnetism plus the
electronic and lattice contributions. In fact, these results plus
the apparent lack of response of the Yb ion in magnetic
measurements (figure~\ref{ybmag}) raised the question of whether
the Yb ion was magnetic at all, despite the indirect indication of
such from the material's lattice parameters \cite{vent92a}.

\begin{figure}[htb]
\begin{center}
\includegraphics[angle=0,width=99mm]{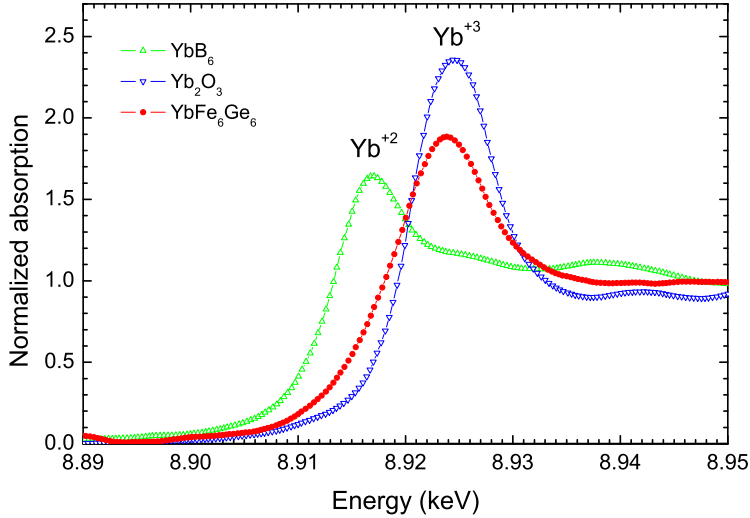}
\caption{\label{xanes} Ytterbium L$_{III}$ X-ray Absorption Near
Edge Structure (XANES) of YbFe$_6$Ge$_6$, compared to those of
Yb$_2$O$_3$ and YbB$_6$.}
\end{center}
\end{figure}

To remove this suspicion, we measured the L$_{III}$ XANES spectrum
of crushed YbFe$_6$Ge$_6$ crystals, presented in
figure~\ref{xanes} together with those of Yb$_2$O$_3$ and YbB$_6$,
used as trivalent \cite{li94a} and divalent \cite{ogi03a}
references respectively. Despite the reduced peak size of
YbFe$_6$Ge$_6$ compared to Yb$_2$O$_3$, which could be attributed
to differences in the metallic vs. ionic environment, it is clear
from the peak position that the Yb ion is essentially trivalent in
YbFe$_6$Ge$_6$, so whatever magnetic response it has is indeed
being masked by the dominant response of the Fe ions.

\section{Discussion}

The anisotropic magnetization experiments performed on our single
crystals clearly establish the occurrence of a low-temperature
anomaly in the magnetic behaviour of YbFe$_6$Ge$_6$, and our
collective set of experimental data gives support to the idea that
the anomalous behaviour results from a sudden reorientation of the
Fe spins, as previously suggested by neutron diffraction and
M\"{o}ssbauer experiments on polycrystals \cite{maz00a}. The
destabilization of the simple axial AFM configuration at lower
temperatures is almost certainly the precursor phenomenon which
allows the otherwise rigidly aligned Fe spins to become more
susceptible to other interactions. However, this destabilization
occurs in FeGe and in most of its derived RFe$_6$Ge$_6$ compounds,
whereas the anomaly has only been documented for R~=~Yb in this
series. So the natural question that follows is: what is unique
about YbFe$_6$Ge$_6$ that may be causing this behaviour?

Given that Yb is an ion with unstable $4f$ shell, it is very
tempting to initially suspect that $4f$-conduction electron
hybridization is somehow related to the unique behaviour, and this
was in fact one of our main motivations to pursue this matter,
since a compound featuring the coexistence of hybridized $4f$
states with ordered $3d$ moments can be potentially very
interesting and rich in phenomena to explore. However, our heat
capacity measurements showed no significant sign of effective
electronic mass enhancement in comparison with LuFe$_6$Ge$_6$, and
resistivity shows no sign of enhancements due to Kondo scattering
- both of which are natural consequences of hybridized $f-d$
states. This leads us to conclude that, at ambient pressure, the
Yb sub-lattice is at best no more than a local-moment paramagnet,
at least down to 1.8~K, and the material's magnetic behaviour is
dominated by the Fe spins.

A second possible approach to attempt an explanation for the
anomaly would be to invoke a structural disorder-induced origin of
the effect, given the easiness with which part of the Yb ions may
be occupying $1a$ sites as shown in figure~\ref{structure}.
However, once again we are faced with the fact that such disorders
are known among all members of the RFe$_6$Ge$_6$ and, if anything,
the measured YbFe$_6$Ge$_6$ sample is less disordered than other
crystals we have grown in the family, since it has the highest
value of $RRR$ we have obtained so far.

A natural consequence of the disorder-induced scenario is that the
effect should be significantly dependent on sample preparation and
thermal history. Indeed we could observe relevant differences in
the measured transitions of earlier batches, including
intermediate steps along the transition, indicative of
intermediate spin configurations (or inhomogeneous samples). In
fact, these characteristics (plus the different field behaviour
for the two measured orientations) allow our results to be
reconciled with the observations of Mazet and Malaman, that the
spin reorientation in their polycrystalline sample began at 85~K
and only about 20\% of the Fe spins deviated from the $c$-axis at
4.2~K. Unfortunately, there were no intentional, systematic
parameter modifications between our batches that could allow
further insight into the relationship between growth conditions
and final magnetic behavior. Such a systematic study would be an
interesting future work, as well as the effects of post-growth
annealing on the crystals.

From all the experimental evidence available, it seems that the
origin of the drastic spin reorientation effect may actually
result from the conjunction of several factors, namely: 1) the
destabilization of the axial Fe spin arrangement upon cooling; 2)
a small but non-negligible Yb-Fe interaction of peculiar RKKY or
some other origin; 3) a crystal field anisotropy of the Yb ions
which competes with the Fe easy axis arrangement; and 4) the
presence of disordered sites which may act as catalysts to
initiate a cascade effect among Fe spins that are strongly coupled
but subject to competing forces.

\section{Conclusion}

The successful growth of single crystals of YbFe$_6$Ge$_6$ and
LuFe$_6$Ge$_6$ has allowed us to investigate their magnetic
anisotropy in the antiferromagetically ordered state, and directly
observe a sudden change in magnetic behaviour of YbFe$_6$Ge$_6$ at
about 63 K, due to a drastic reorientation of the Fe spins, an
unusual effect for the RFe$_6$Ge$_6$ series where the Fe and R
sub-lattices are known to behave quite independently. Our work has
answered some of the questions raised by previous investigations
which showed evidence of such an effect to occur in this compound,
and eliminated Yb $4f$-conduction electron hybridization as a
likely candidate to explain its occurrence. However, the actual
source of the effect, and especially the nature and strength of
the Yb-Fe interaction which is almost certainly the crucial
element behind this unique behaviour, remains an open question to
be investigated in further detail through more powerful
experimental techniques on the now available crystals. The
possibility of inducing hybridization in the Yb electronic levels
and following the interactions between these and the ordered Fe
spins may be a worthwhile endeavor, as well as mapping of how
thermal history affects disorder and magnetic behaviour in this
compound.

\ack

The crystal growths and low-temperature measurements were
partially performed both at Ames Laboratory, Iowa State University
and at the Materials Science Center, N-BARD, Hiroshima University.
Thanks to F. Iga for supplying the YbB$_6$ sample reference
compound. This work was supported by a Grant-in-Aid for Scientific
Research (COE Research 13CE2002) of MEXT Japan, and by the
Director for Energy research, Office of Basic Energy Sciences,
Department of Energy USA. Ames Laboratory is operated for the US
Department of Energy by Iowa State University under Contract No.
W-7405-Eng-82. XANES measurements were performed with the approval
of SPring-8/JASRI (Proposal No. 2005A0628-NXa-np).

\end{document}